\documentclass[aps,prl,twocolumn]{revtex4}

\usepackage{graphicx}

\usepackage{bm}
\usepackage{amsmath}
\usepackage{amssymb}

\newcommand{\ocite}{\onlinecite}

\newcommand{\x}{\text}
\newcommand{\pd}{\partial}

\newcommand{\dg}{\dagger}

\newcommand{\lt}{\left}
\newcommand{\rt}{\right}

\newcommand{\f}{\frac}
\newcommand{\tf}{\tfrac}

\newcommand{\sq}{\sqrt}
\newcommand{\lbl}{\label}

\newcommand{\cd}{\cdot}

\newcommand{\p}{\perp}

\newcommand{\nm}{\hat{0}}
\newcommand{\um}{\hat{1}}

\newcommand{\tm}{\times}

\newcommand{\eq}[1]{Eq.~(\ref{eq:#1})}
\newcommand{\eqs}[2]{Eqs.~(\ref{eq:#1}) and (\ref{eq:#2})}

\newcommand{\eqn}[1]{(\ref{eq:#1})}
\newcommand{\eqsn}[2]{(\ref{eq:#1}) and (\ref{eq:#2})}

\newcommand{\figr}[1]{Fig.~\ref{fig:#1}}

\newcommand{\spc}{\mbox{ }}

\newcommand{\beq}{\begin{equation}}
\newcommand{\eeq}{\end{equation}}
\newcommand{\beqar}{\begin{eqnarray}}
\newcommand{\eeqar}{\end{eqnarray}}
\newcommand{\beqarn}{\begin{eqnarray*}}
\newcommand{\eeqarn}{\end{eqnarray*}}
\newcommand{\ba}{\begin{array}}
\newcommand{\ea}{\end{array}}
\newcommand{\bwt}{\begin{widetext}}
\newcommand{\ewt}{\end{widetext}}

\newcommand{\sgn}{{\rm sgn}\,}

\newcommand{\rarr}{\rightarrow}

\newcommand{\Og}{\mathbb{O}}

\newcommand{\Tg}{\mathbb{T}}

\newcommand{\ex}{{\text e}}

\newcommand{\ix}{{\text i}}

\newcommand{\Lx}{{\text L}}

\newcommand{\Ox}{{\text O}}

\newcommand{\Tx}{{\text T}}

\newcommand{\ph}{\hat{p}}

\newcommand{\Hh}{\hat{H}}

\newcommand{\Jh}{\hat{J}}

\newcommand{\Mh}{\hat{M}}

\newcommand{\psih}{\hat{\psi}}

\newcommand{\Ec}{\mathcal{E}}

\newcommand{\nv}{{\bf 0}}

\newcommand{\pb}{{\bf p}}

\newcommand{\Jb}{{\bf J}}

\newcommand{\Jbh}{\hat{\Jb}}

\newcommand{\al}{\alpha}
\newcommand{\be}{\beta}

\newcommand{\Ga}{\Gamma}

\newcommand{\la}{\lambda}
\newcommand{\La}{\Lambda}

\newcommand{\eps}{\varepsilon}
\newcommand{\e}{\epsilon}

\begin{document}
\title{Universality and stability of the edge states of chiral-symmetric topological semimetals
and surface states of the Luttinger semimetal}
\author{Maxim Kharitonov, Julian-Benedikt Mayer, and Ewelina M. Hankiewicz}
\address{Institute for Theoretical Physics and Astrophysics,
University of W\"urzburg, 97074 W\"urzburg, Germany}
\begin{abstract}

We theoretically demonstrate that the chiral structure of the nodes of
nodal semimetals is responsible for the existence and universal local properties of the edge states in the vicinity of the nodes.
We perform a general analysis of the edge states
for an isolated node of a 2D semimetal,
protected by {\em chiral symmetry}
and characterized by the topological winding number $N$.
We derive the asymptotic chiral-symmetric boundary conditions
and find that there are $N+1$ universal classes of them.
The class determines the numbers of flat-band edge states
on either side off the node in the 1D spectrum
and the winding number $N$ gives the {\em total} number of edge states.
We then show that the edge states of chiral nodal semimetals are {\em robust}:
they persist in a finite-size {\em stability region} of parameters of chiral-asymmetric terms.
This significantly extends the notion of 2D and 3D topological nodal semimetals.
We demonstrate that the Luttinger model with a quadratic node for $j=\frac32$ electrons
is a 3D topological semimetal in this new sense and predict that $\alpha$-Sn, HgTe, possibly Pr$_2$Ir$_2$O$_7$,
and many other semimetals described by it are topological and exhibit surface states.
\end{abstract}
\maketitle

\begin{figure}
\includegraphics[width=.47\textwidth]{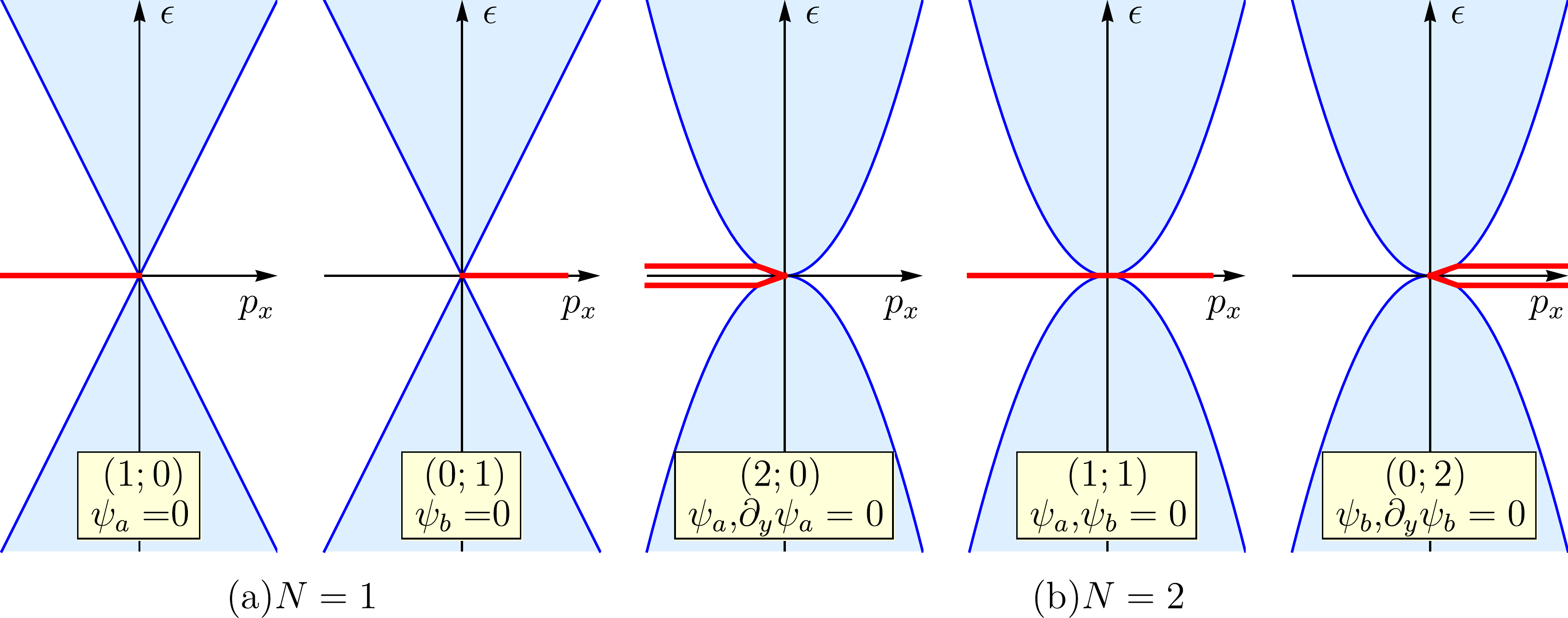}
\caption{
Universal local structure of the edge states of 2D chiral-symmetric nodal semimetals,
with the Hamiltonian $\Hh_N$ [\eq{H}] and boundary conditions \eqn{cbc} of classes $(N_a,N_b)$,
illustrated for (a) $N=1$ and (b) $N=2$.
There are $N_{a,b}$ flat edge-state bands (red) on the two sides off the node
and their total number $N_a+N_b=N$ is equal to the winding number;
for $(2,0)$ and $(0,2)$, the degenerate bands are split for visibility.
}
\lbl{fig:chsymm}
\end{figure}

{\em Introduction.}
Edge states in 2D nodal semimetals
have been demonstrated in numerous theoretical calculations~\cite{Fujita,Nakada,Brey,Akhmerov,FZhang,Castro,Arikawa,MartinBlanter,Xu,LiTao,Kunstmann,Bahamon,Jaskolski,
DeshpandeWinkler,Chiu,ChiuRMP},
mainly for models describing monolayer and bilayer graphene.
Their existence is attributed~\cite{Chiu,ChiuRMP}
to the topological invariants characterizing the nodes, the winding numbers $N$,
which are well-defined in the presence of {\em chiral symmetry}.
Still, up to now, the general structure of the edge states of chiral-symmetric 2D nodal semimetals
and its relation to the winding numbers have not yet been explicitly established for arbitrary $N$.

In this Letter, we carry out this task locally in the Brillouin zone (BZ),
by performing a general analytical analysis of the edge states in the vicinity of an isolated node.
As the main advancement, we derive the most general form
of the asymptotic boundary conditions (BCs) that respect chiral symmetry.
We find that there are $N+1$ discrete universal classes of them.
These classes describe all possible universal structures of the edge states (\figr{chsymm})
and establish their connection to the winding number $N$.

We then address the stability properties of the edge states
and show that they are robust under the effects of chiral symmetry breaking.
This allows for a significant extension of the notion of a topological nodal semimetal in both 2D and 3D.
As an important application of the developed framework,
we demonstrate that the Luttinger model~\cite{L}
with a quadratic node for $j=\f32$ electrons,
describing materials like $\al$-Sn~\cite{BP,W,Madelung}, HgTe~\cite{BP,W,Madelung}, and $\x{Pr}_2\x{Ir}_2\Ox_7$~\cite{Pr},
exhibits surface states and is a 3D topological semimetal in this new more general sense.

{\em 2D chiral-symmetric nodal semimetal.}
First, we consider an isolated chiral node of a 2D semimetal arising
at some point in the BZ from a degeneracy of two electron levels, to be denoted $a$ and $b$.
We assume that the Hamiltonian for the
two-component wave function
    $\psih=(\psi_a, \psi_b)^\Tx$
has the form
\beq
    \Hh_N(p_x,p_y)=\lt(\ba{cc} 0& p_-^N \\ p_+^N & 0 \ea\rt),
    \spc p_\pm=p_x\pm\ix p_y,
\lbl{eq:H}
\eeq
to the leading order in momentum $(p_x,p_y)$ in the vicinity of the node;
$N$ is a positive integer.

The Hamiltonian has chiral symmetry~\cite{ChiuRMP}:
under the transformation
\beq
    \psih\rarr \tau_z\psih, \spc \tau_z=\x{diag}(1,-1),
\lbl{eq:cspsi}
\eeq
it changes its sign,
$
    \tau_z\Hh_N(p_x,p_y)\tau_z^\dg=-\Hh_N(p_x,p_y)
$.
Due to chiral symmetry,
$\Hh_N$ is characterized by a well-defined topological invariant,
the winding number $N$,~\cite{ChiuRMP} related to the Berry phase $\pi N$.

We stress that the winding number $N$ is a {\em local} topological characteristic of the node in the BZ.
The properties of the edge states that we study are also local,
and we do not address the reasons
for the existence of the node.
We assume that the node is isolated from possible other nodes at different points in the BZ,
also in the presence of the edge~\cite{isolatednode}.

{\em Universal asymptotic chiral-symmetric boundary conditions.}
We first derive the most general form of the BCs for the Hamiltonian $\Hh_N$
that satisfy chiral symmetry.
This derivation is free of any microscopic assumptions
and, beside chiral symmetry, invokes only two natural requirements:
long-wavelength limit and vanishing of the probability current perpendicular to the boundary.

We assume the sample occupies the half-plane $y>0$.
Since the Schr\"odinger equation $\Hh_N(\ph_x,\ph_y)\psih=\e\psih$ ($\ph_{x,y}=-\ix\pd_{x,y}$)
is a differential equation of order $N$ in each component $\psi_{a,b}$,
BCs at the boundary $y=0$ are a set of $N$ linear homogeneous
(meaning that the linear combinations are equated to zero)
relations for the derivatives $\pd_y^n\psi_{a,b}\equiv\pd^n_y\psi_{a,b}(x,y=0)$, $n=0,\ldots,N-1$
($\pd^0_y\psi_{a,b}=\psi_{a,b}$ being the components themselves);
we drop the arguments of the functions in the BCs formulas for brevity.

The long-wavelength limit means the following.
Any linear relation involving derivatives of different order necessarily contains spatial scales.
Consider, for example, a relation
$
    \pd_y\psi_a+l\pd_y^2\psi_a=0,
$
characterized by a spatial scale $l$.
In the long-wavelength limit, at spatial scales larger than $l$,
the second term $l\pd_y^2\psi_a$ becomes negligible and the relation reduces to $\pd_y\psi_a=0$.
Thus, in the BCs satisfying the requirement of the long-wavelength limit,
to be referred to as {\em asymptotic} BCs,
only the derivatives of the same order can be present {\em in one relation}.
Therefore, for a given order $n$, there is {\em either no} BCs, {\em or one} BC
\beq
    c_{an}\pd^n_y\psi_a+c_{bn}\pd^n_y\psi_b=0
\lbl{eq:bcn}
\eeq
with dimensionless coefficients $c_{an,bn}$, {\em or two} BCs
\beq
    \pd_y^n\psi_a=0 \x{ and }\pd_y^n\psi_b=0
\lbl{eq:bcn2}
\eeq
with both derivatives vanishing individually.

Demanding chiral symmetry, we find that BC \eqn{bcn} remains invariant under the
transformation \eqn{cspsi} only if one of the coefficients $c_{an,bn}$ is zero,
so that BC \eqn{bcn} reduces to either
\[
    \pd_y^n\psi_a=0 \x{ or } \pd_y^n\psi_b=0.
\]
Combined with the possibility \eqn{bcn2},
we find that under chiral symmetry the most general form of the asymptotic BCs
is when some $N$ out of $2N$ derivatives
are individually nullified:
\beq
    \pd_y^n\psi_\la=0, \spc (\la,n)\in\La.
\lbl{eq:cbc0}
\eeq
Here, $\la=a,b$ and $n=0,\ldots,N-1$, and $\La$ is a subset of size $N$
of $2N$ indices $(\la,n)$ labelling the said derivatives.

These $C^N_{2N}$ types of BCs can be sorted into $N+1$ groups $(N_a,N_b)$ with $N_{a,b}=0,...,N$,
such that
\beq
    N_a+N_b=N,
\lbl{eq:N}
\eeq
according to the numbers $N_{a,b}$ of constraints
imposed on the derivatives $\pd^n_y\psi_{a,b}$ of a given component.

Finally, the hermiticity of the full Hamiltonian
demands that the probability current perpendicular to the boundary must vanish at the boundary
\beq
    j_y(x,y=0)=0.
\lbl{eq:jy=0}
\eeq
The expressions for the probability current for $\Hh_N$ [\eq{H}] read
[see Supplemental Material (SM)~\cite{SM}]
\beq
    j_y=-\ix(j_+-j_+^*),\spc
    j_+=\sum_{n=0}^{N-1}(\ph_+^{N-1-n}\psi_a)^* \ph_-^n\psi_b.
\lbl{eq:j}
\eeq
The bilinear form \eqn{jy=0} must vanish identically for any $\psih$.
Inspecting \eq{j}~\cite{SM},
we find that, for a given group $(N_a,N_b)$,
{\em only one} BC among \eqn{cbc0} is allowed,
the one with the {\em lowest-order} derivatives nullified:
\beq
    \psi_a,\ldots,\pd^{N_a-1}_y\psi_a,\psi_b,\ldots,\pd_y^{N_b-1}\psi_b=0.
\lbl{eq:cbc}
\eeq
These are all possible asymptotic chiral-symmetric current-conserving BCs for the chiral-symmetric Hamiltonian $\Hh_N$ [\eq{H}].
There are $N+1$ {\em classes}
$(N_a,N_b)$ of them. This is the first key result of this work.
For $N=1,2$, all BC classes $(N_a,N_b)$ are shown in \figr{chsymm}.

{\em Edge states and the winding number $N$.}
Further, the edge states for the Hamiltonian $\Hh_N$ [\eq{H}] and BCs \eqn{cbc} can be found explicitly~\cite{SM}.
Taking the plane-wave form
$
    \psih(x,y)=\psih(p_x,y)\ex^{\ix p_x x}
$
with momentum $p_x$ along the edge,
we find $N_b$ edge-state solutions
\[
    \psih_n(p_x>0,y)=
    (1,0)^\Tx
    y^n \ex^{-p_x y},\spc n=N_a,\ldots,N-1,
\]
at $p_x>0$ and $N_a$ edge-state solutions
\[
    \psih_n(p_x<0,y)=
    (0,1)^\Tx
    y^n\ex^{+p_x y},\spc n=N_b,\ldots,N-1,
\]
at $p_x<0$.
All solutions have zero energy $\e=0$ and thus represent {\em flat bands}.

Thus, we have shown that for a 2D chiral-symmetric nodal semimetal
with both the bulk Hamiltonian [\eq{H}] and BCs [\eq{cbc}] obeying chiral symmetry,
a set of flat-band edge states always exists asymptotically in the vicinity of an {\em isolated} node.
The sum of the numbers $N_{a,b}$ of the edge-state bands
on both sides $p_x\gtrless 0$ off the node in the 1D edge
spectrum is equal to the winding number $N$, \eq{N} and
\figr{chsymm}. This is the second key result of this work~\cite{graphene}.

And so, the {\em total} number of the edge-state bands
is determined solely by the local in the BZ bulk characteristic of the node, the winding number $N$,
irrespective of the chiral BC class $(N_a,N_b)$,
which determines the numbers of the edge-state bands on either side off the node.
The specific class $(N_a,N_b)$ is in general determined by the bulk Hamiltonian also away from the node,
as well as by the orientation and microscopic structure of the edge.
Still, $(N_a,N_b)$ are also topological numbers,
since they cannot be changed by continuously changing the system parameters or surface orientation
{\em while preserving the chiral symmetry}.
Changes in the numbers $(N_a,N_b)$ of the edge-state bands
can occur by changing the surface orientation
only when projections of different nodes onto the surface collapse,
which necessarily requires the presence of more than one node in the BZ.
The BC class $(N/2,N/2)$, however, possible for even $N$,
could be realized for just a single node in the whole BZ;
the BC class $(1,1)$ will be relevant below.

\begin{figure}
\includegraphics[width=.47\textwidth]{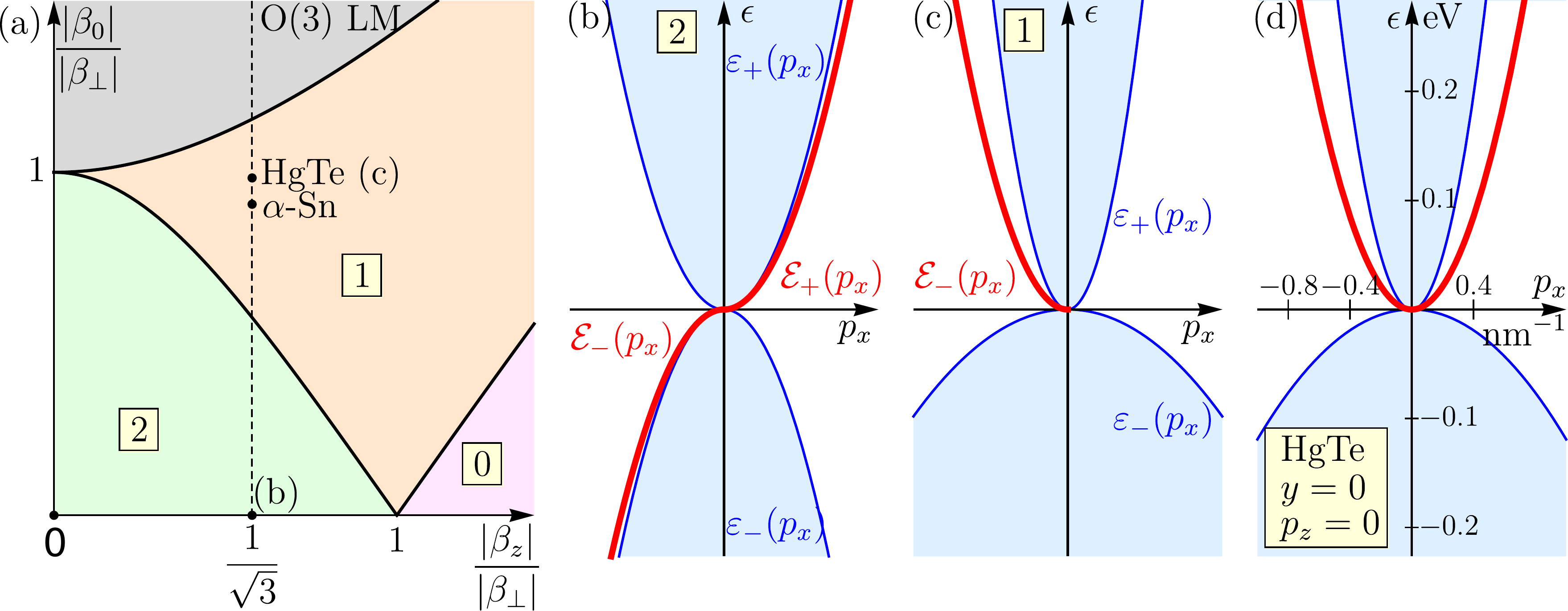}
\caption{
(a),(b),(c) Edge states of the chiral-asymmetric quadratic Hamiltonian $\Hh_2^\be$ [\eq{H2b}]
with chiral-symmetric BCs \eqn{cbc11} of class $(1,1)$:
(a) Edge-state stability phase diagram in the parameter plane $(\be_z,\be_0)$;
dashed line $|\be_z|/|\be_\p|=1/\sq3$ corresponds to the $\Ox(3)$-symmetric Luttinger model;
(b), (c)
Edge states \eqn{Ec2b} (red)
for $(\be_z,\be_0)/|\be_\p|=\f1{\sq3}(1,0)$
and $\f1{\sq3}(1,1.7)$
in the stability regions 2
and 1 with two [$\Ec_\pm(p_x)$] and one [$\Ec_-(p_x)$] bands, respectively.
(d) Surface states (red) of the Luttinger model $\Hh^\Lx(\pb)$ [\eq{HL}] with BCs \eqn{LMbc}
for $y>0$ sample at $p_z=0$ for the parameters
of HgTe with neglected inversion asymmetry.
}
\lbl{fig:H2}
\end{figure}

{\em Stability of chiral-symmetric edge states under breaking of chiral symmetry.}
The found edge states of a 2D chiral-symmetric nodal semimetal [\eqs{H}{cbc}]
are stable under the effects of chiral-symmetry breaking.
Both the bulk Hamiltonian and BCs may contain terms that break chiral symmetry,
making them deviate from their chiral-symmetric forms \eqsn{H}{cbc}.
As chiral-asymmetric terms are introduced,
edge states can disappear only by merging with particle or hole continua of bulk states.
Since for preserved chiral symmetry the flat edge-state bands are positioned at $\e=0$,
it takes finite strength of chiral-asymmetric terms to force edge states merge with bulk states.
We thus introduce the notion of the {\em stability region} of chiral-symmetric
edge states: it is a finite-size region in the parameter space of chiral-asymmetric terms around
the point of chiral symmetry, within which edge states persist.
This is the third key result of this work.

We illustrate the effect of the chiral asymmetry of the bulk Hamiltonian for $N=2$
by considering the model
\beq
    \Hh_2^\be(p_x,p_y)=\lt(\ba{cc} (\be_0+\be_z)p_+p_- & \be_\p p_-^2 \\ \be_\p p_+^2 & (\be_0-\be_z)p_+p_- \ea\rt)
\lbl{eq:H2b}
\eeq
with complex $\be_\p$ and real $\be_{0,z}$.
The terms due to $\be_{0,z}$ break chiral symmetry;
at $\be_{0,z}=0$, $\Hh_2^\be$ reduces to the chiral-symmetric form $\Hh_2$ [\eq{H}].
The bulk spectrum of $\Hh_2^\be$ is
$
    \eps_\pm(p_\p)=(\be_0\pm\sq{|\be_\p|^2+\be_z^2})p_\p^2, \spc p_\p^2=p_x^2+p_y^2.
$
For $\be_0^2<|\be_\p|^2+\be_z^2$,
the system is in the nodal semimetal regime we will focus on,
with particle and hole bands.

We calculate the edge states for chiral-symmetric BCs
\beq
    \psi_a,\psi_b=0
\lbl{eq:cbc11}
\eeq
of class $(1,1)$, \figr{chsymm}(b).
We obtain~\cite{SM}
the edge-state dispersion relations
\beq
    \Ec_\pm(p_x)=2|\be_\p|\f{\be_0|\be_\p|\pm\be_z\sq{|\be_\p|^2+\be_z^2-\be_0^2}}{|\be_\p|^2+\be_z^2} p_x^2,
\lbl{eq:Ec2b}
\eeq
at $p_x\gtrless 0$, respectively, \figr{H2}(a),(b),(c).
In the plane $(\be_z,\be_0)$ of chiral-asymmetry parameters, \figr{H2}(a),
the semimetal region $|\be_0|<\sq{|\be_\p|^2+\be_z^2}$ consists of three subregions
2,1,0, labelled according to the numbers of the edge-state bands:
stability region 2 (green), which contains the point of chiral symmetry $\be_{0,z}=0$,
and in which both bands $\Ec_\pm(p_x)$ at $p_x\gtrless0$,
originating from the chiral-symmetric edge states of the BC class $(1,1)$ [\figr{chsymm}(b)], persist,
\figr{H2}(b);
stability region 1 (orange), where only one of the bands $\Ec_\pm(p_x)$
on one side off $p_x=0$ exists, \figr{H2}(c);
and region 0 (magenta), where the edge states are absent.
Regions 2 and 1 and regions 1 and 0
are separated by the curves $|\be_0|=\pm\f{|\be_\p|^2-\be_z^2}{\sq{|\be_\p|^2+\be_z^2}}$, respectively.

We illustrate the effect of chiral asymmetry of the BCs for the linear node ($N=1$) in SM~\cite{SM}.

{\em Extended notion of 2D and 3D topological nodal semimetals.}
The above findings offer a significant extension
of the notion of a 2D topological nodal semimetal:
one may regard a 2D nodal semimetal as topological if it belongs to the stability region
of some chiral-symmetric 2D nodal semimetal.
The edge states of the latter are ensured by a well-defined topological invariant, the winding number $N$.
Yet, exact chiral symmetry is not required,
and the edge states will persist in the former
as long as chiral-symmetric terms are dominant.
The above example, \eqs{H2b}{cbc11}, \figr{H2}, is a 2D topological nodal semimetal in this sense.
This definition is then readily extended to 3D:
one may regard a 3D nodal semimetal as topological, if its 2D reductions
to at least some planes in momentum space passing through the node(s)
are 2D topological nodal semimetals in the above sense.
In this case, the 3D nodal semimetal will exhibit surface states of topological origin~\cite{anisotropy}.

This viewpoint has wide-reaching implications,
since chiral terms are ubiquitous in the Hamiltonians of 2D and 3D nodal semimetals,
though exact chiral symmetry is not necessarily present.
It allows one to prove the existence and topological origin of the edge or surface states
in semimetal systems by relating them to 2D chiral-symmetric models,
even in cases when a precise topological invariant may be hard or impossible to define.

{\em Luttinger model for $j=\f32$ electrons as a 3D topological semimetal.}
One important example of a 3D semimetal with a quadratic node that may be regarded as topological
in this new sense is the 4-band Luttinger model~\cite{L} (LM)
for electrons with $j=\f32$ angular momentum (Luttinger semimetal):
\beq
    \Hh^\Lx(\pb)=(\al_0+\tf52\al_z)\pb^2\um_4-2\al_z(\Jbh\cd\pb)^2+\al_\square\Mh_\square(\pb).
\lbl{eq:HL}
\eeq
Here, $\um_4$ is the unit matrix of order 4, and $\Jbh=(\Jh_x,\Jh_y,\Jh_z)$ are the angular-momentum matrices.

It describes the local electron band structure around the $\Ga$ point of a material
with full cubic point group $\Og_h$ with inversion and time-reversal symmetry.
It is the most general form up to quadratic order in momentum $\pb=(p_x,p_y,p_z)$
allowed by these symmetries. All four $j=\f32$ states are degenerate at $\pb=\nv$ due to $\Og_h$ symmetry.
Odd-$\pb$ terms are prohibited by inversion.
The terms $\pb^2\um_4$ and $(\Jbh\cd\pb)^2$ are invariants of the full spherical symmetry group $\Ox(3)$ with inversion.
Their linear combination $\Hh^\Lx(\pb)|_{\al_\square=0}$,
characterized by two parameters $\al_{0,z}$, is the $\Ox(3)$-symmetric LM;
its bulk spectrum has two double-degenerate bands
$\eps^{\Lx,\Ox(3)}_\pm(p)=(\al_0\pm2|\al_z|)p^2$, $p=|\pb|$;
$|\al_0|<2|\al_z|$ is the nodal semimetal regime.
The additional term $\Mh_\square(\pb)=\Jh_x^2p_x^2+\Jh_y^2p_y^2+\Jh_z^2p_z^2-\f25(\Jbh\cd\pb)^2-\f15\Jbh^2\pb^2$
with parameter $\al_\square$ is a cubic anisotropy term, which
arises from lowering the symmetry from spherical to cubic, $\Ox(3)\rarr\Og_h$.

The LM Hamiltonian $\Hh^\Lx(\pb)$ must be supplemented by proper physical BCs.
We find~\cite{SM} that the asymptotic BCs for the wave function
$\psih^\Lx=(\psi^\Lx_{+\f32},\psi^\Lx_{+\f12},\psi^\Lx_{-\f12},\psi^\Lx_{-\f32})^\Tx$
(subscripts indicate $j_z$) of the LM following from the 6-band Kane model with hard-wall BCs,
describing an interface with vacuum, have the form
\beq
    \psih^\Lx=\nm.
\lbl{eq:LMbc}
\eeq
The Kane model describes materials like $\al$-Sn and HgTe (see below).

At $p_z=0$, $\Hh^\Lx(p_x,p_y,0)$ is block-diagonal:
the pairs $(\psi^\Lx_{+\f32},\psi^\Lx_{-\f12})$ and $(\psi^\Lx_{+\f12},\psi^\Lx_{-\f32})$ of states decouple;
the BCs \eqn{LMbc} lead to the chiral-symmetric BCs \eqn{cbc11} of class $(1,1)$ for each pair.
For $\Ox(3)$ symmetry, the respective $2\tm 2$ blocks of $\Hh^\Lx(p_x,p_y,0)|_{\al_\square=0}$
are of the form $\Hh_2^\be(p_x,p_y)$ [\eq{H2b}] of opposite chiralities,
with parameters $\be_{0,z}=\al_{0,z}$ and $\be_\p=-\sq3\al_z$.
The LM is thus on the line $|\be_z|/|\be_\p|=1/\sq3$
in the parameter plane $(\be_0,\be_z)$ of $\Hh_2^\be$, \figr{H2}(a),
and always belongs to the stability regions 2 or 1,
as determined by the ratio $|\al_0|/|\al_z|$;
$|\al_0|=|\al_z|$ is the transition point between regions 2 and 1.
For $y>0$ sample, the surface-state dispersion relations for $(\psi^\Lx_{+\f32},\psi^\Lx_{-\f12})$
are $\Ec_\pm^{\Lx,\Ox(3)}(p_x,p_z=0)=\f{\sq3}2(\sq3\al_0\pm\sgn\al_z\sq{4\al_z^2-\al_0^2})p_x^2$ at $p_x\gtrless0$, respectively [\eq{Ec2b}],
shown in \figr{H2}(b) and \figr{H2}(c)
for $\al_0=0$ and the parameters $\al_0/\al_z=1.7$ of HgTe (see below)
belonging to regions 2 and 1, respectively. When combined with the dispersion relations
for $(\psi^\Lx_{+\f12},\psi^\Lx_{-\f32})$, $\Ec_\mp^{\Lx,\Ox(3)}(p_x,p_z=0)$
at $p_x\gtrless0$, respectively, full surface-state bands of the LM are obtained.

Since for spherical symmetry the same
holds for any other orientation of the momentum plane,
we conclude that the $\Ox(3)$-symmetric 3D LM
exhibits 2D surface states in the whole nodal semimetal regime $|\al_0|<2|\al_z|$,
two bands for $|\al_0|<|\al_z|$ and one band for $|\al_z|<|\al_0|<2|\al_z|$,
for any orientation of the surface and any direction of 2D momentum along the surface.
According to the above stability arguments,
this result also holds upon including the cubic-anisotropy term $\al_\square\Mh_\square(\pb)$,
as long as $\al_\square$ is small enough.
We have thus proven that the $\Og_h$-symmetric LM $\Hh^\Lx(\pb)$ [\eq{HL}] with the BCs \eqn{LMbc}
is a 3D topological nodal semimetal in the sense of this work.
This is the fourth key result of this work.

Our results on the LM are relevant to a multitude of real materials
either with exact cubic symmetry $\Og_h$ or in which deviations from it are small.
Among materials with $\Og_h$ that exhibit a quadratic node is $\al$-Sn~\cite{BP,W,Madelung};
a prime example of a material with weakly broken $\Og_h$
is HgTe~\cite{BP,W,Madelung} with a tetrahedral point group $\Tg_d$.
The LM parameters for $\al$-Sn and HgTe with neglected inversion asymmetry,
extracted from Ref.~\ocite{Madelung}, are
$(\al_0,\al_z)=(9.31,5.94)/m_e$
and $(\al_0,\al_z,\al_\square)=(7.28,4.29,-0.44)/m_e$, respectively, where $m_e$ is the electron mass.
For $\Ox(3)$ symmetry, they both belong to region 1, as indicated in \figr{H2}(a),
and thus exhibit one band of 2D surface states.
\figr{H2}(d) shows the surface states for $y>0$ sample at $p_z=0$
for the parameters of HgTe including the cubic anisotropy $\al_\square$.
Recently, a quadratic node was predicted and likely observed in
$\x{Pr}_2\x{Ir}_2\Ox_7$~\cite{Pr};
according to our findings, one or two bands of surface states can be anticipated
for this material, although a separate analysis would be desirable.
We thus predict that $\al$-Sn, HgTe,
and many other semimetal materials described by the LM are topological in the sense of this work.

Deviations from cubic symmetry $\Og_h$ due to
breaking of inversion, rotational (strain, confinement), or time-reversal (magnetism)
symmetries modify the low-energy band structure,
causing the quadratic node of the LM to gap out or split into linear nodes.
A variety of resulting topological phases,
such as a topological insulator~\cite{BrueneHgTe3DTI,BaumHgTe3DTI,BrueneHgTe3DTIb},
Weyl semimetal~\cite{Ruan},
and quantum anomalous Hall insulator~\cite{Liu},
has been predicted or observed.
According to our findings, the $\Og_h$-symmetric quadratic nodal semimetal of the LM [\eqs{HL}{LMbc}]
can be regarded as the {\em parent, highest-symmetry topological phase} for these phases,
with its own surface states of topological origin.

{\em Future directions.}
Relations between the topological properties of these phases
is an interesting future direction.
Among other possible applications and extensions of this work are:
relation of the local in the BZ properties of the edge and surface states established here to their global properties,
such as those of 3D Weyl semimetals~\cite{ChiuRMP};
addressing the edge states in graphene and similar 2D systems~\cite{Fujita,Nakada,Brey,Akhmerov,FZhang,Castro,Arikawa,MartinBlanter,Xu,LiTao,Kunstmann,Bahamon,Jaskolski,
DeshpandeWinkler,Chiu,ChiuRMP}
within this framework~\cite{graphene};
the role of electron interactions for the edge and surface states
in quadratic nodal semimetals in 2D (bilayer graphene) and 3D (LM),
where interactions are predicted~\cite{Sun,Lemonik,Cvetkovic,Murray,Abrikosov,Balents,Herbut,Goswami}
to result in interesting physics in the bulk.

{\em Acknowledgements.}
We appreciate discussions with B.~Trauzettel and S.~Juergens.
We acknowledge financial  support  from  the  DFG  via  SFB  1170
``ToCoTronics'' and the ENB Graduate School on Topological Insulators.


\begin{thebibliography}{99}
\bibitem{isolatednode}
Coupling between the nodes becomes possible if they project onto the same 1D momentum.
This happens, for example, for armchair graphene edge.
This could also happen for other graphene edge orientations
for edges ``decorated'' with certain periodic structures
that partially lower translational symmetry,
so that the momenta of the nodes fold onto the same 1D momentum.
Our formalism allows for straightforward generalization to the cases of multiple coupled nodes.
\bibitem{SM}
See Supplemetal Material at [URL],
which includes Refs.~\ocite{BP,W,Madelung,LLIII,vOstaay,Rhim},
for the derivations of the boundary conditions, edge states,
and Luttinger model.
\bibitem{anisotropy}
For systems with strong enough spatial anisotropy,
one can envision a regime,
where the edge/surface states exist only in a finite range of edge orientations in 2D
and of surface orientations and momentum directions along the surface in 3D.
The edge/surface states would disappear by merging with the bulk states upon changing these parameters.
\end{thebibliography}
\end{document}


\title{Supplemental Material: Universality and stability of the edge states of chiral-symmetric topological semimetals
and surface states of the Luttinger semimetal}
\author{Maxim Kharitonov, Julian-Benedikt Mayer, and Ewelina M. Hankiewicz}
\address{Institute for Theoretical Physics and Astrophysics,
University of W\"urzburg, 97074 W\"urzburg, Germany}

\maketitle

\setcounter{equation}{0}
\renewcommand\theequation{S\arabic{equation}}

\setcounter{figure}{0}
\renewcommand\thefigure{S\arabic{figure}}


\section{Probability current}

Here, we derive the probability current
\[
    \jb=(j_x,j_y)
\]
for the chiral-symmetric Hamiltonian $\Hh_N$
[Eq.~(1)].
The procedure is standard and analogous to the one for the conventional quadratic Hamiltonian~\cite{LLIIIS}.
The probability density
\beq
    \rho=\psih^\dg\psih=\psi_a^*\psi_a+\psi_b^*\psi_b
\lbl{eq:rho}
\eeq
of a wavefunction $\psih=(\psi_a,\psi_b)^\Tx$ satisfying the time-dependent Schr\"odinger equation
\beq
    \ix\pd_t\psih=\Hh_N\psih
\LRa \lt\{\ba{c}
    \ix\pd_t\psi_a=\ph_-^N\psi_b,\\
    \ix\pd_t\psi_b=\ph_+^N\psi_a
    \ea\rt.
\lbl{eq:seq}
\eeq
must satisfy the continuity equation
\[
    \pd_t\rho+\n\cd\jb=0, \spc \n=(\pd_x,\pd_y).
\]
In the integral form
\beq
    \int_{\Vc}\dx\rb\,\pd_t\rho+\int_{\pd\Vc}\dx\s\,\jb=0,
\lbl{eq:contint}
\eeq
the rate of change of the probability in a 2D space region $\Vc$
must be compensated by the current flow through the boundary $\pd\Vc$
of the region.
Substituting the expressions \eqn{seq} for $\pd_t\psi_{a,b}$ into $\pd_t\rho$
and integrating $\int_{\Vc}\dx\rb\,\pd_t\rho$
in parts sufficient number of times, we arrive at
\beq
    j_x=j_++j_-,\spc j_y=-\ix(j_+-j_-),\spc j_-=j_+^*,
\lbl{eq:jSM}
\eeq
with $j_+$ given in
Eq.~(8).

In particular,
\beq
    N=1:\spc
    j_+=\psi_a^*\psi_b,
\lbl{eq:jN1}
\eeq
\[
    N=2:\spc
    j_+=\psi_a^*\ph_-\psi_b+(\ph_+\psi_a)^*\psi_b.
\]

The formulas
(8) and \eqn{jSM}
can naturally be understood as follows.
For a plane-wave function $\psih\sim\ex^{\ix(p_x x+p_y y)}$ the current is given by the derivatives of the Hamiltonian over momentum,
\[
    j_+=\f{\pd \Hh_{N,ab}}{\pd p_-}=Np_-^{N-1}
,\spc
    j_-=\f{\pd \Hh_{N,ba}}{\pd p_+}=Np_+^{N-1},
\]
and
Eqs.~(8) and \eqn{jSM}
represent the properly symmetrized operator version of this.

\section{Boundary conditions, current conservation constraints}


Here, we prove in more detail how the current conservation constraint
(7)
restricts the allowed asymptotic chiral-symmetric BCs from the most general form
(5)
to the final form
(9).

Only such form of BCs is allowed that, for any wave function $\psih=(\psi_a,\psi_b)^\Tx$ satisfying them,
the current component $j_y(x,0)=0$
[Eq.~(7)]
perpendicular to the edge vanishes identically at the edge $y=0$.
The current component $j_+(x,0)$ in $j_y$
[Eq.~(8)]
is a sum of the terms
\beq
    \pd_x^{n_{ax}}\pd_y^{n_{ay}}\psi_a^*(x,0)\,\pd_x^{n_{bx}}\pd_y^{n_{by}}\psi_b(x,0)
\lbl{eq:psiterms}
\eeq
with nonnegative integers $n_{ax,ay,bx,by}=0,\ldots,N-1$ such that
\[
    n_{ax}+n_{ay}+n_{bx}+n_{by}=N-1
\]
and $j_-(x,0)=j_+^*(x,0)$ is the sum of the corresponding conjugate terms.
Since for chiral symmetry the only allowed forms
(5)
of BCs
are when some individual derivatives $\pd^n_y\psi_{a,b}(x,0)$ vanish
(while linear relations
(3)
between different components are prohibited),
the current $j_y(x,0)$ can be nullified only if {\em all} the terms \eqn{psiterms}
vanish {\em individually}.

The terms \eqn{psiterms} involve derivatives both perpendicular to ($\pd_y$) and along ($\pd_x$) the edge.
Since these terms must vanish at {\em any} point $(x,0)$ along the edge,
this is equivalent to vanishing individually of the terms
\beq
    \pd_y^{n_{ay}}\psi_a^*(x,0)\,\pd_y^{n_{by}}\psi_b(x,0)
\lbl{eq:dypsi}
\eeq
identically for all $x$ for all  $n_{ay,by}=0,\ldots,N-1$ such that
\[
    n_{ay}+n_{by}\leq N-1.
\]

This is possible for chiral-symmetric BCs
(5)
only if, for given $(N_a,N_b)$,  the {\em lowest-order} derivatives are nullified
at the edge $y=0$, 
as expressed in
Eq.~(9).

\section{Edge states for chiral symmetry, details}

Here, we provide details of the derivation of the edge states for chiral-symmetric
Hamiltonian
(1)
and BCs
(9).
Taking the plane-wave form
\[
    \psih(x,y)=\psih(p_x,y)\ex^{\ix p_x x}
\]
with momentum $p_x$ along the edge,
we first look for the general solution to the Sch\"odinger equation
\[
    \Hh_N(p_x,\ph_y)\psih(p_x,y)=\e\psih(p_x,y).
\]
We find the edge states at energy $\e=0$ and have checked that there are no other edge states at $\e\neq0$.
At $\e=0$, the components are decoupled and we get the equations
\[
    (p_x+\pd_y)^N\psi_a(p_x,y)=0,
\spc    (p_x-\pd_y)^N\psi_b(p_x,y)=0.
\]

There are $N$ independent solutions for each component:
\[
    \psi_{an}(p_x,y)=y^n\ex^{-p_x y},
\spc
    \psi_{bn}(p_x,y)=y^n\ex^{+p_x y},
\spc n=0,\ldots,N-1.
\]
At $p_x>0$, $\psi_{an,bn}(p_x,y)$ decay and grow into the bulk, $y\rarr+\iy$, respectively,
and so, only the solutions with finite $\psi_a(p_x,y)$ and vanishing $\psi_b(p_x,y)\equiv0$ components are allowed.
Applying the chiral BCs
(9),
we get that
$N_a$ boundary conditions for $\psi_a(p_x,y)$
yield $N-N_a=N_b$ independent edge-state solutions $\psih_n(p_x>0,y)$, $n=N_a,\ldots,N-1$,
with $\e=0$, provided in the Main Text.
Similarly, at $p_x<0$, we find $N-N_b=N_a$ edge-state solutions $\psih_n(p_x<0,y)$, $n=N_b,\ldots,N-1$,
with $\e=0$, provided in the Main Text.

\section{Edge states for chiral-asymmetric Hamiltonian $\Hh_2^\be(p_x,p_y)$, details}

Here, we provide details of the derivation of the edge states for chiral-asymmetric
quadratic Hamiltonian $\Hh_2^\be(p_x,p_y)$
[Eq.~(10)]
and chiral-symmetric BCs
(11).
Taking the plane-wave form
\[
    \psih(x,y)=\psih(p_x,y)\ex^{\ix p_x x}
\]
with momentum $p_x$ along the edge,
we first look for the general solution to the Sch\"odinger equation
\beq
    \Hh_2^\be(p_x,\ph_y)\psih(p_x,y)=\e\psih(p_x,y).
\lbl{eq:H2beq}
\eeq
Its characteristic equation
\[
    \det[\Hh_2^\be(p_x,p_y)-\e\um_2]=0
\]
has four momentum solutions
\[
    p_y=\pm\ix\sq{p_x^2-\f{\e}{\be_\pm}},
\]
where $\be_\pm=\be_0\pm\sq{|\be_\p|^2+\be_z^2}$
are the curvatures of the electron and hole bulk bands $\eps_\pm(p_\p)=\be_\pm p_\p^2$;
we consider the nodal semimetal regime, where $\be_+>0$ and $\be_-<0$,
and $\e$ such that $p_x^2>\f{\e}{\be_\pm}$.

The partial solutions to \eq{H2beq} corresponding to
the pair $p_y=\ix\sq{p_x^2-\f{\e}{\be_\pm}}$
of momentum solutions  are
\[
    \psih_\pm(p_x,\e)\ex^{-\sq{p_x^2-\f{\e}{\be_\pm}}y},
    \spc
    \psih_\pm(p_x,\e)=\lt(\ba{c}
        2p_x^2-\f{\e}{\be_\pm}+2p_x\sq{p_x^2-\f{\e}{\be_\pm}}
        \\
        \f{\e}{\be_\p}(1-\f{\be_0+\be_z}{\be_\pm})
    \ea\rt).
\]
For the sample at $y>0$,
these solutions decay into the bulk and are admitted,
while the partial solutions with $p_y=-\ix\sq{p_x^2-\f{\e}{\be_\pm}}$
grow into the bulk and are prohibited.

Applying the BCs
(11)
to the linear combination
\[
    \psih(p_x,y)=C_+\psih_+(p_x,\e)\ex^{-\sq{p_x^2-\f{\e}{\be_+}}y}+C_-\psih_-(p_x,\e)\ex^{-\sq{p_x^2-\f{\e}{\be_-}}y}
\]
of the decaying solutions, we find
that a nontrivial solution with nonzero $C_\pm$ exists when
\[
    \psi_{+a}(p_x,\e)\psi_{-b}(p_x,\e)-\psi_{+b}(p_x,\e)\psi_{-a}(p_x,\e)=0.
\]
Solving this equation with respect to $\e$,
we obtain the edge-state dispersion relations $\Ec_\pm(p_x)$
[Eq.~(12)]
at $p_x\gtrless 0$,
respectively,
and find the phase diagram in the plane $(\be_0,\be_z)$ of chiral-asymmetry parameters, presented in
Fig.~2(a).

\section{Edge states for chiral-asymmetric boundary condition}

\begin{figure}
\includegraphics[width=.35\textwidth]{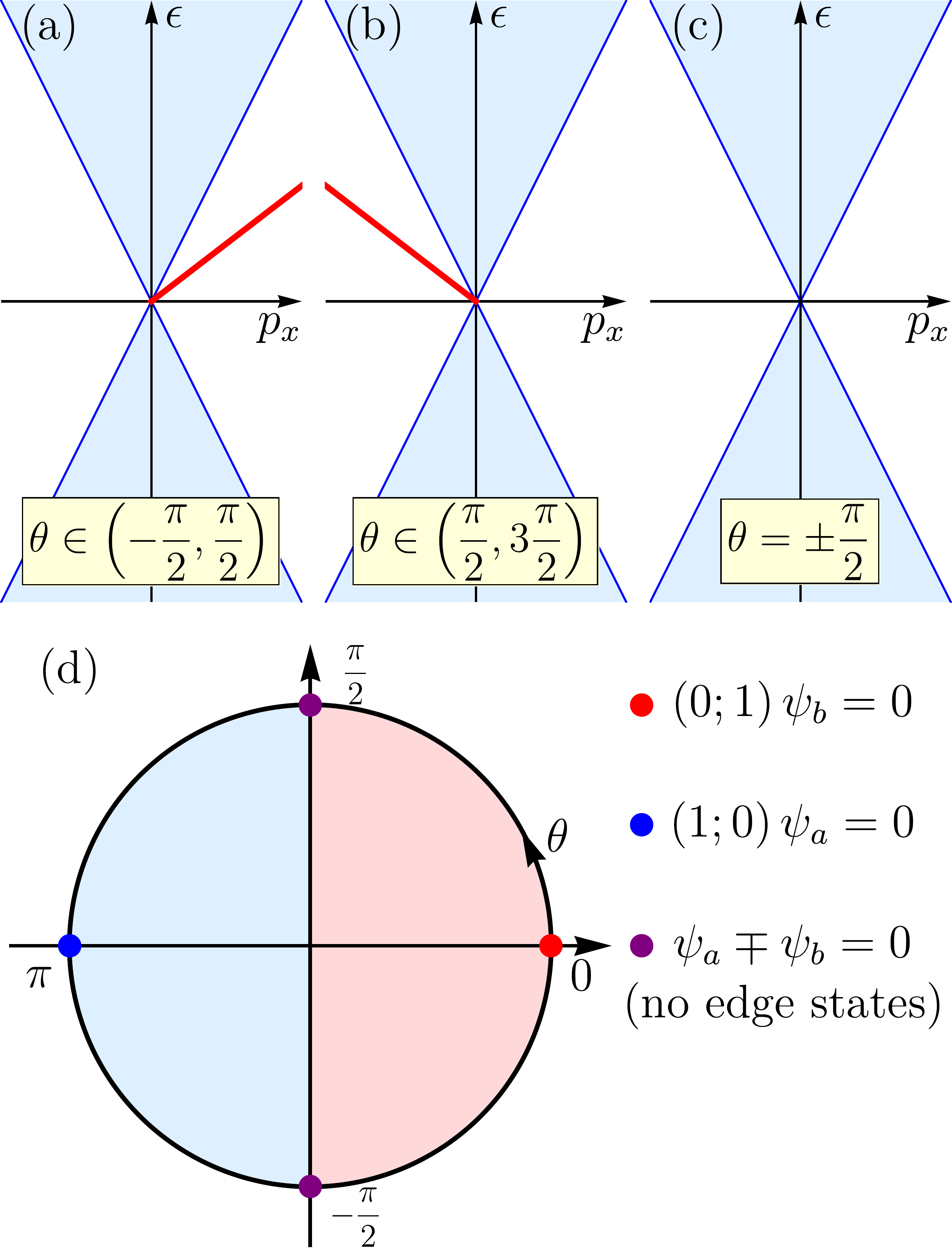}
\caption{
(a),(b),(c) Edge states $\Ec(p_x)$ [\eq{Ec1}] (red) for the linear-node chiral-symmetric Hamiltonian $\Hh_1(p_x,p_y)$ [$N=1$, Eq.~(1)]
and chiral-asymmetric BC \eqn{cabcN1}; compare to the case of chiral-symmetric BCs in Fig.~1(a).
(d) The circle of the angle parameter $\tht$ of the most general, chiral-asymmetric BC \eqn{cabcN1} for $\Hh_1(p_x,p_y)$.
}
\lbl{fig:N1chasymm}
\end{figure}

Here, we illustrate the effect of chiral asymmetry of BCs for the linear-node chiral-symmetric  Hamiltonian $\Hh_1(p_x,p_y)$ [$N=1$, Eq.~(1)].
We derive the most general form of the BC for this Hamiltonian~\cite{Babak}.
Mathematically, such BC
is a single linear homogeneous relation
between the two components of the wave function;
without the loss of generality, it can be written as
\beq
    \ex^{\ix\f{\phi}2}\sin\tf\tht2\psi_a-\ex^{-\ix\f{\phi}2}\cos\tf\tht2\psi_b=0,
\lbl{eq:cabcN10}
\eeq
parameterized by two real angles $\tht$ and $\phi$.

For a terminated system, the BC must only satisfy the fundamental constraint that the current perpendicular to the boundary vanishes.
This means that, for $y=0$ edge, $j_y=-\ix(\psi_a^*\psi_b-\psi_b^*\psi_a)=0$ must vanish identically for any wave function satisfying \eq{cabcN10}.
This is possible only if $\phi=0$.
And so, the most general form of the BC for the linear-node chiral-symmetric Hamiltonian $\Hh_1(p_x,p_y)$ [$N=1$, Eq.~(1)]
reads
\beq
    \sin\tf\tht2\psi_a-\cos\tf\tht2\psi_b=0.
\lbl{eq:cabcN1}
\eeq
All unique forms of the BC are parameterized by the angle $\tht$ covering the {\em full circle}, \figr{N1chasymm}(d).


It is then straightforward to find~\cite{Babak} the edge states for $\Hh_1(p_x,p_y)$ [$N=1$, Eq.~(1)] with the BC \eqn{cabcN1}.
Taking the plane-wave form
\[
    \psih(x,y)=\psih(p_x,y)\ex^{\ix p_x x}
\]
with momentum $p_x$ along the edge,
we first look for the general solution to the Sch\"odinger equation
\beq
    \Hh_1(p_x,\ph_y)\psih(p_x,y)=\e\psih(p_x,y).
\lbl{eq:H2beq}
\eeq
Its characteristic equation
\[
    \det[\Hh_1(p_x,p_y)-\e\um_2]=0
\]
has two momentum solutions
\[
    p_y=\pm\ix\sq{p_x^2-\e^2}, \spc |\e|<|p_x|.
\]
For $y>0$ sample, only $p_y=\ix\sq{p_x^2-\e^2}$ corresponds
to the decaying wave function
\beq
    \psih(p_x,y)
    =\lt(\ba{c}
        p_x+\sq{p_x^2-\e^2}
        \\
        \e
    \ea\rt)
    \ex^{-\sq{p_x^2-\e^2}y}
\lbl{eq:psi1}
\eeq
and is admitted.
Applying the BC \eqn{cabcN1} to \eq{psi1},
we get the equation
\[
    \sin\tf\tht2(p_x+\sq{p_x^2-\e^2})-\cos\tf\tht2\e=0.
\]
Solving it with respect to $\e$,
we find one branch of edge states on one side off the node,
at $p_x>0$ for $\tht\in\Tht_a=(-\f\pi2,\f\pi2)$ and at $p_x<0$ for $\tht\in\Tht_b=(\f\pi2,\f{3\pi}2)$, \figr{N1chasymm}(a) and (b).
In both cases, the edge-state dispersion relation reads
\beq
    \Ec(p_x)=p_x \sin\tht.
\lbl{eq:Ec1}
\eeq
The sectors $\Tht_a=(-\f\pi2,\f\pi2)$ and $\Tht_b=(\f\pi2,\f{3\pi}2)$
contain the points of chiral symmetry $\tht=0$
[BC class $(0,1)$: $\psi_b=0$, red dot in \figr{N1chasymm}(d)] and $\tht=\pi$ [BC class $(1,0)$: $\psi_a=0$, blue dot in \figr{N1chasymm}(d)],
respectively, see Fig.~1(a), and thus represent the {\em stability regions}, as defined in the Main Text,
in which the respective chiral-symmetric edge states persist.
As $\tht$ deviates from one of the chiral-symmetric points $\tht=0,\pi$,
the edge states deviate from $\e=0$ acquiring a finite velocity $\pd_{p_x}\Ec(p_x)=\sin\tht$.
The edge states disappear by merging with the bulk bands only upon reaching the points $\tht=\pm\f\pi2$
[purple dots in \figr{N1chasymm}(d)].
The stability regions $\Tht_{a,b}$ are thus separated only by two points $\tht=\pm\f\pi2$
and the edge states persist even for significant deviations of the BC from chiral symmetry.


The chiral-asymmetric BC \eqn{cabcN1} with general $\tht$ and the corresponding edge states \eqn{Ec1}
apply~\cite{vOstaay}, for instance, to a realistic model of graphene for the zigzag edge in the vicinity of the nodes (valleys).
Its chiral-symmetric limits $\psi_b=0$ or $\psi_a=0$ with flat edge states, Fig~1(a),
apply to the chiral-symmetric lattice model under the microscopic assumption of nearest-neighbor hopping only.
Upon including next-nearest-neighbor hopping,
these BCs transform to the chiral-asymmetric form \eqn{cabcN1},
and the edge-state dispersion relation acquires finite slope, \eq{Ec1}.

\subsection{The case of absent edge states}

The edge states thus exist for all values of $\tht$ in the BC \eqn{cabcN1}
except $\tht=\pm\f{\pi}2$, when the BC has one of the forms
\beq
    \psi_a\mp\psi_b=0,
\lbl{eq:cabcN1abs}
\eeq
respectively.
We point out that even though these cases are realized only
at points
in the 1D parameter space (circle of $\tht$) of possible chiral-asymmetric BCs,
it does not necessarily mean that such cases are negligibly rare in the space of possible
models.
These cases
can be dictated by the
microscopic structure of the model.
Below we present two such examples.

\subsubsection{Infinite mass model}

As the first example,
suppose that the region $y>0$ is described by the gapless Hamiltonian $\Hh_1(p_x,p_y)$ [Eq.~(1)]
and the region $y<0$ is described by the Hamiltonian
\beq
    \Hh_{1\De}(p_x,p_y)=\Hh_1(p_x,p_y)+\De\tau_z
\lbl{eq:H1De}
\eeq
with an additional gap term $\De\tau_z$ with $\De>0$.

At energies $|\e|<\De$, the wave functions decay into the region $y<0$.
At small energies $|\e|\ll\De$, the decay spatial scale $1/\De$ [note that the velocity is set to unity in $\Hh_1(p_x,p_y)$] is much shorter
than the typical scale $1/|\e|$ of variation of the wave function in the region $y>0$,
and the model may be substituted by an equivalent one,
in which the wave function is nonvanishing only in the gapless region $y>0$
and satisfies an effective BC at the interface $y=0$.

To derive the BC,
to the leading order in $\e/\De\ll 1$,
it is sufficient to find the general solution at $\e=0$ that decays into the region $y<0$.
In the gapless region $y>0$, the general solution is
an arbitrary coordinate-indepedent spinor
\[
    \psih(x,y>0)=\lt(\ba{c} \psi_a\\\psi_b \ea\rt);
\]
%
in the gapped region $y<0$, it is a decaying function
\[
    \psih(x,y<0)=C\lt(\ba{c} 1\\ 1\ea\rt)\ex^{\De y}.
\]
Connecting these functions continuously at $y=0$ and excluding the coefficient $C$,
we get the relation
\[
    \psi_a-\psi_b=0
\]
between the wave-function components in the gapless region $y>0$.
For a wave function $\psih(x,y)$ varying at scales larger than $1/\De$,
this relation becomes the BC at $y=0$.
This is the BC of the type \eqn{cabcN1abs} with $\tht=\f{\pi}2$,
for which the edge states are absent, \figr{N1chasymm}(c).

Regarding symmetry,
the term $\De\tau_z$ in \eq{H1De} breaks chiral symmetry,
and since the gap $\De$ is infinite relative to the energy $\e$ of interest
(for this reason, this BC is often called the ``infinite mass'' BC),
chiral symmetry is broken so strongly
that the edge states are completely absent.

\subsubsection{Lattice model with chiral symmetry broken by the edge}

\begin{figure}
\includegraphics[width=.45\textwidth]{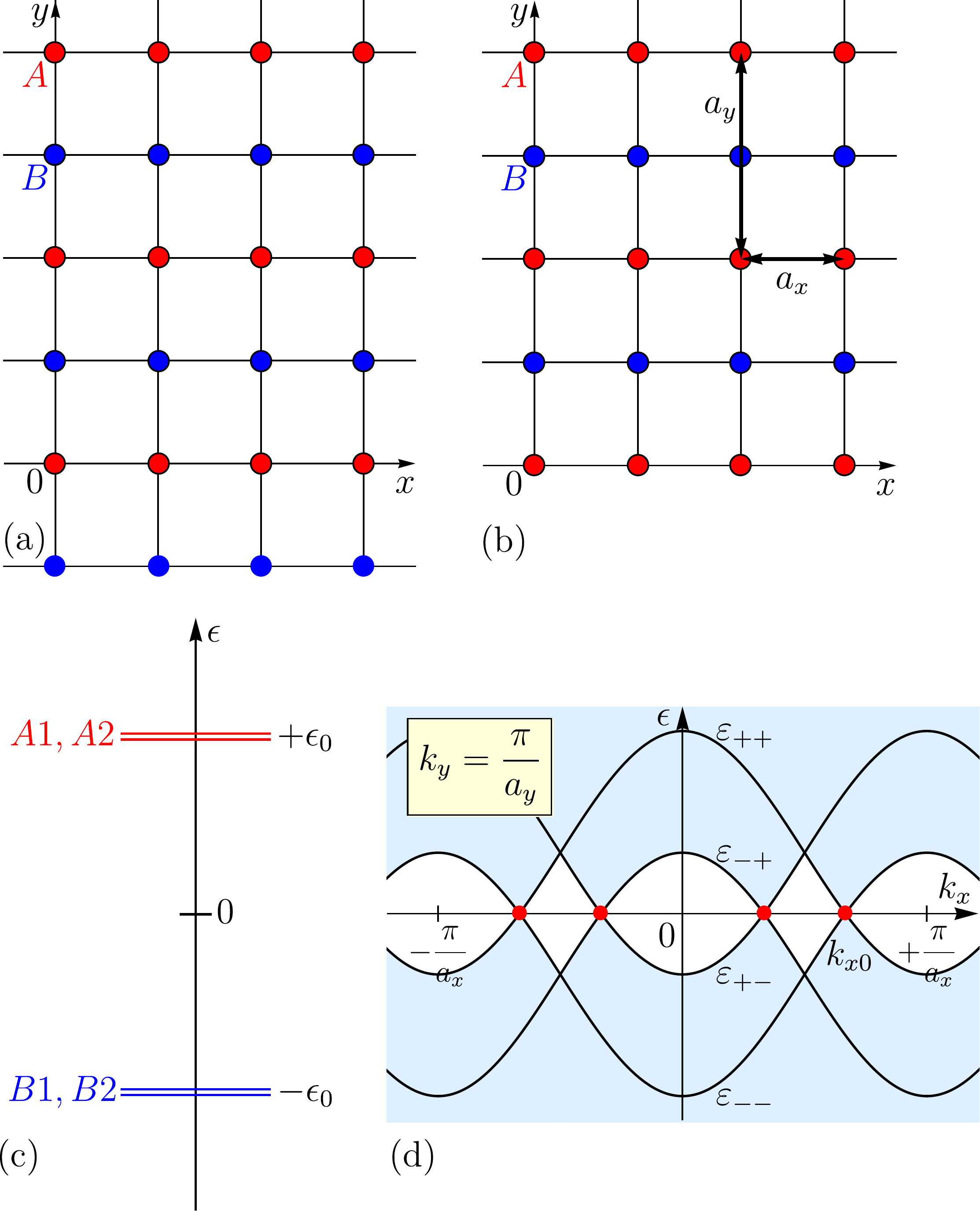}
\caption{
The lattice model studied in Sec. V B of Ref.~\ocite{Rhim}.
(a),(b) Lattice structure with two cases of the edge along the $x$ direction.
(c) On-site energy levels.
(d) Electron spectrum $\eps_{\pm\pm}=\eps_{\pm\pm}(k_x,k_y=\f{\pi}{a_y})$ [\eq{epsky=pi}] at $k_y=\f{\pi}{a_y}$.
The shaded light-blue region is the continuum of the bulk states as all $k_y$ are spanned.
}
\lbl{fig:modelB}
\end{figure}

Another example is the lattice model studied in Sec. V B of Ref.~\ocite{Rhim}, see \figr{modelB}.
The model consists of two sublattices, $A$ and $B$, and two orbitals, 1 and 2, at each site, \figr{modelB}(a) and (b);
the lattice wave function is a four-component spinor
\[
    \Psih(\rb)=\lt(\ba{c} \Psi_{A1}(\rb) \\ \Psi_{A2}(\rb) \\ \Psi_{B1}(\rb) \\ \Psi_{B2}(\rb) \ea\rt),
\]
where
\[
    \rb=\ab_x n_x+\ab_y n_y
\]
is the discrete radius vector on the primitive rectangular lattice, with periods $\ab_x=(a_x,0)$ and $\ab_y=(0,a_y)$
and $n_{x,y}$ taking integer values.

The tight-binding Hamiltonian in the momentum space reads
\beq
    \Hch(k_x,k_y)=\e_0 T_{z0}+2t_x\cos(k_x a_x)  T_{0x}+2 t_y \cos(\tf{k_y a_y}2) T_{xz}.
\lbl{eq:HcB}
\eeq
Here, $T_{\al\be}=\tau_\al^{AB}\ot\tau_\be^{12}$, $\al,\be=0,x,y,z$,
are the basis matrices in the direct product of the sublattice ($AB$) and orbital ($12$) spaces
($\tau_{0,x,y,z}$ denote the unity and Pauli matrices in the respective spaces);
$\pm\e_0$ are the energies of both orbitals 1 and 2 at $A$ and $B$ sites, respectively, \figr{modelB}(c);
$t_{x}$ is the nearest-neighbor hopping amplitude in the $x$ direction
(hopping between 1 and 2 orbitals of the same sublattice);
$t_y$ is the nearest-neighbor hopping amplitude in the $y$ direction
(hopping between the same orbitals of different sublattices,
with different signs of the amplitude).
We consider a slightly more general model than in Ref.~\ocite{Rhim},
with unrelated $t_{x,y}$ and $a_{x,y}$ for $x$ and $y$ directions.

The spectrum of the Hamiltonian \eqn{HcB} consists of four bands
\[
    \eps(k_x,k_y)=\pm\sq{[\e_0\pm 2 t_x \cos(k_x a_x)]^2+[2 t_y\cos(\tf{k_y a_y}2)]^2}.
\]
At $k_y=\f{\pi}{a_y}$, there is partial decoupling of the basis states;
the spectrum takes the form
\beq
    \eps_{\pm\pm}(k_x,k_y=\tf{\pi}{a_y})=\pm \e_0 \pm 2 t_x\cos(k_x a_x),
\lbl{eq:epsky=pi}
\eeq
\figr{modelB}(d),
and the eigenstates at every $k_x$, labeled respectively, are
\[
    \Psih_{++}=\f1{\sq2}\lt(\ba{c} +1\\+1\\0\\0\ea\rt)
,\spc
    \Psih_{+-}=\f1{\sq2}\lt(\ba{c} +1\\-1\\0\\0\ea\rt)
,\spc
    \Psih_{-+}=\f1{\sq2}\lt(\ba{c} 0\\0\\+1\\+1\ea\rt)
,\spc
    \Psih_{--}=\f1{\sq2}\lt(\ba{c} 0\\0\\+1\\-1\ea\rt).
\]

There are four linear nodes at $\e=0$ on the line $k_y=\f{\pi}{a_y}$, at $k_x=\pm\f1{a_x}\arccos\lt(\pm\f{\e_0}{2 t_x}\rt)$.

We consider the edge along the $x$ direction,
with two options of the last row being either the $B$ [\figr{modelB}(a)] or $A$ [\figr{modelB}(b)] sublattice,
and derive the effective low-energy Hamiltonian and BC for one of the nodes,
performing the $k\cd p$ expansion.
We consider the node at $\kb_0=(k_{x0},\f{\pi}{a_y})$,
$k_{x0}=\f1{a_x}\arccos\lt(-\f{\e_0}{2t_x}\rt)$, originating from the band crossing of $\Psih_{++}$ and $\Psih_{--}$ states.

In the vicinity of the node,
the low-energy degrees of freedom
are described by the two-component spinor
\beq
    \psith(\rb)=\lt(\ba{c} \psit_a(\rb)\\ \psit_b(\rb)\ea\rt),
\lbl{eq:psiB}
\eeq
slowly varying at the lattice scale $a_{x,y}$, in the lattice wave function of the form
\beq
    \Psih(\rb)=[\psit_a(\rb)\Psih_{++}+\psit_b(\rb)\Psih_{--}]\ex^{\ix\kb_0\rb}.
\lbl{eq:PsiB}
\eeq

We obtain the linearized Hamiltonian
\beq
    \Hth(p_x,p_y)
    =v_x p_x\tau_z+ v_y p_y\tau_x,
\lbl{eq:HtB}
\eeq
for $\psith(\rb)$,
where $(p_x,p_y)$ is the deviation from the node momentum $\kb_0$
and $(v_x,v_y)=-(2t_x a_x\sin k_{x0}, t_y a_y )$.


The lattice BCs for the edge along the $x$ direction
read
\[
    \Psi_{A1}(n_x,n_y=-1)=\Psi_{A2}(n_x,n_y=-1)=0
\]
for \figr{modelB}(a), and
\[
    \Psi_{B1}(n_x,n_y=-1)=\Psi_{B2}(n_x,n_y=-1)=0
\]
for \figr{modelB}(b).
Plugging the wave function \eqn{PsiB} into them, we obtain the BCs
\beq
    \psit_a(x,y=0)=0
\lbl{eq:psita}
\eeq
and
\beq
    \psit_b(x,y=0)=0,
\lbl{eq:psitb}
\eeq
respectively,
for the wave function \eqn{psiB} of the low-energy model.


It can be shown that the linearized Hamiltonian \eqn{HtB} with either \eqn{psita} or \eqn{psitb} BC has no edge states.
In Ref.~\ocite{Rhim}, it was found
that the edge states are absent for the edge along the $x$ direction
for the initial lattice model \eqn{HcB} at all momenta $k_x$.

This behavior is in full agreement with our findings.
Upon the change of basis
\[
    \psith=\Uh\psih
,\spc
    \Uh=\f1{\sq2}\lt(\ba{cc} \ex^{\ix\f{\pi}4} & -\ex^{\ix\f{\pi}4}\\ \ex^{-\ix\f{\pi}4} & \ex^{-\ix\f{\pi}4} \ea\rt),
\]
the Hamiltonian
\[
    \Hh(p_x,p_y)=U^\dg\Hth(p_x,p_y) U=-v_x p_x\tau_x+v_y p_y \tau_y
\]
for
$\psih=(\psi_a,\psi_b)^T$ takes the form of $\Hh_1(p_x,p_y)$ and the BCs \eqsn{psita}{psitb}
take the form of \eq{cabcN1abs},
which are
indeed the only cases ($\tht=\pm\f{\pi}2$) of the chiral-asymmetric BC \eqn{cabcN1}, for which the edge states are absent.

Regarding symmetry,
we notice
that the bulk lattice Hamiltonian \eqn{HcB} possesses chiral symmetry:
\[
    \Sh\Hch(k_x,k_y)\Sh^\dg=-\Hch(k_x,k_y)
\]
for the chiral-symmetry operator
\[
    \Sh=T_{xy} \mbox{ or } T_{yz},
\]
or their arbitrary unitary linear combination.
The linear nodes at $k_y=\f{\pi}{a_y}$ therefore have well-defined winding numbers $N=1$ (of both signs),
with the chiral symmetry operator $\tau_z$ in the $\psih$ basis.

Under such chiral-symmetry operation, the sublattices $A$ and $B$ are interchanged.
This interchange becomes impossible if the lattice has an edge along the $x$ direction, \figr{modelB}(a) and (b).
And so, chiral symmetry is broken by the edge in this model,
even though it is preserved in the bulk.
This results in the chiral-asymmetric BCs in both the lattice and low-energy models
and
is the ultimate reason for the absence of the edge states.


\section{Luttinger model from Kane model}

\subsection{6-band Kane model}

For studying the surface states of the 4-band LM for $j=\f32$ states,
its bulk Hamiltonian $\Hh^\Lx(\pb)$
[Eq.~(13)]
must be supplemented by proper physical BCs.
In this paper, we derive the asymptotic BCs for the LM
that follow from a more general Kane model~\cite{BPS,WS} (KM) with ``hard-wall'' BCs.
The 6-band KM includes, in addition to $j=\f32$ quartet,
a $j=\f12$ doublet of {\em opposite} inversion parity
and describes a large family of semiconductor materials~\cite{BPS,WS,MadelungS},
in which $j=\f32$ states originate from a $p$ orbital in the presence of spin-orbit interactions
and $j=\f12$ states originate from an $s$ orbital.
Considering the KM is instructive from a more general standpoint,
for the purpose of demonstrating a systematic ``folding'' procedure,
where the high-energy $j=\f12$ states of a larger Hilbert space of the KM are
consistently eliminated to generate the effective bulk Hamiltonian
and BCs of the LM with the smaller Hilbert space that contains only the low-energy $j=\f32$ states.

So, the Hamiltonian and the wave function of the KM have the general block structure
\beq
    \Hh^\Kx(\pb)=\lt(\ba{cc} \Hh_{\f12\f12}(\pb) & \Hh_{\f12\f32}(\pb) \\ \Hh_{\f32\f12}(\pb) & \Hh_{\f32\f32}(\pb) \ea \rt),
    \spc \pb=(p_x,p_y,p_z),
\lbl{eq:HK}
\eeq
\beq
    \Psih
    =\lt(\ba{c}\Psih_{\f12} \\ \Psih_{\f32} \ea\rt)
    ,\spc
    \Psih_{\f12}=\lt(\ba{c}\Psi_{\f12,+\f12} \\ \Psi_{\f12,-\f12} \ea\rt)
    ,\spc
    \Psih_{\f32}=\lt(\ba{c}\Psi_{\f32,+\f32} \\ \Psi_{\f32,+\f12} \\ \Psi_{\f32,-\f12} \\ \Psi_{\f32,-\f32} \ea\rt),
\lbl{eq:Psi}
\eeq
in the space of $j=\f12$ and $j=\f32$ states; here, $j_z=\pm\f12,\pm\f32$ denote
the angular momentum projections on the $z$ axis.

Like the LM, the KM describes the local electron band structure around the $\Ga$ point $\pb=\nv$.
For full cubic symmetry $\Og_h$ with inversion and time reversal symmetry,
the most general form up to quadratic order in $\pb$ reads
\[
    \Hh_{\f12\f12}(\pb)=(\De+\ga_{\f12}\pb^2)\um_2,
\]
\[
    \Hh_{\f12\f32}(\pb)
    =v\lt(\ba{cccc}
        -\f1{\sq2}p_+ & \sq{\f23} p_z & \f1{\sq6}p_- & 0\\
        0 & -\f1{\sq6}p_+ & \sq{\f23} p_z & \f1{\sq2} p_-
    \ea\rt)
    ,\spc
    \Hh_{\f32\f12}(\pb)=\Hh_{\f12\f32}^\dg(\pb),
\]
\[
    \Hh_{\f32\f32}(\pb)=\ga_0\pb^2\um_4+\ga_z\Mh(\pb)+\ga_\square\Mh_\square(\pb),
\]
\[
    \Mh(\pb)=\f52\pb^2\um_4-2(\Jbh\cd\pb)^2
    =\lt(\ba{cccc}
        p_+p_--2p_z^2 & -\sq3 2 p_-p_z & -\sq3 p_-^2& 0\\
        -\sq3 2 p_+p_z & -p_+p_-+2p_z^2 & 0 & -\sq3 p_-^2\\
        -\sq3 p_+^2&0& -p_+p_-+2p_z^2 & \sq3 2 p_-p_z\\
        0&-\sq3 p_+^2& \sq3 2 p_+p_z& p_+p_--2p_z^2
    \ea\rt),
\]
\[
    \Mh_\square(\pb)=\Jh_x^2p_x^2+\Jh_y^2p_y^2+\Jh_z^2p_z^2-\f25(\Jbh\cd\pb)^2-\f15\Jbh^2\pb^2,
\]
\[
    \Jbh=(\Jh_x,\Jh_y,\Jh_z),\spc \Jh_\pm=\Jh_x\pm\ix\Jh_y,\spc
    \Jh_+=\lt(\ba{cccc} 0&\sq3&0&0 \\ 0&0&2&0 \\ 0&0&0&\sq3 \\ 0&0&0&0 \ea\rt),
    \spc \Jh_-=\Jh_+^\dg
,\spc
    \Jh_z=\lt(\ba{cccc} +\tf32&0&0&0 \\ 0&+\tf12&0&0 \\ 0&0&-\tf12&0 \\ 0&0&0&-\tf32 \ea\rt)
,\spc
    \Jbh^2=\f{15}4\um_4.
\]
Here and below, $\um_n$ denotes the unit matrix of order $n$.

This form follows from the method of invariants~\cite{BPS,WS} ($k\cd p$ method).
The $j=\f32$ and $j=\f12$ states form a four- and a two-dimensional (projective) irreducible representation of $\Og_h$, respectively.
They correspond to four- and two-fold-degenerate levels at $\pb=\nv$,
which we take to be at energies $\e=0$ and $\e=\De$, respectively.
Due to opposite inversion parities of the $j=\f32$ and $j=\f12$ states,
the cross-product block $\Hh_{\f12\f32}(\pb)$ contains only odd powers of $\pb$,
while the self-product blocks $\Hh_{\f12\f12}(\pb)$ and $\Hh_{\f32\f32}(\pb)$ contain only even powers of $\pb$.
The cross-product block $\Hh_{\f12\f32}(\pb)$ contains one linear-in-$\pb$ invariant,
with the velocity coefficient $v$, which is real due to time-reversal symmetry.
Within the $j=\f12$ states,
the block $\Hh_{\f12\f12}(\pb)$ contains one invariant $\um_2\pb^2$ quadratic in $\pb$.
Within the $j=\f32$ states, the block $\Hh_{\f32\f32}(\pb)$
contains three invariants $\pb^2\um_4$, $(\Jbh\cd\pb)^2$, $\Mh_\square(\pb)$
of $\Og_h$ and time-reversal symmetry quadratic in $\pb$.
Understandably, the block $\Hh_{\f32\f32}(\pb)=\Hh^\Lx(\pb)|_{\al_{0,z,\square}=\ga_{0,z,\square}}$
has the structure of the LM
(13),
since this is the most general form allowed by symmetry.

In fact, all the above invariants of $\Og_h$, except for $\Mh_\square(\pb)$,
are also invariants of the full spherical symmetry group $\Ox(3)$ with inversion.
Therefore, the KM $\Hh^\Kx(\pb)|_{\ga_\square=0}$
without the $\Mh_\square(\pb)$ term is the most general form allowed by $\Ox(3)$ and time-reversal symmetries.
The term $\Mh_\square(\pb)$ thus represents {\em cubic anisotropy},
which arises from lowering the symmetry $\Ox(3)\rarr\Og_h$;
it transforms as a linear combination of the states of angular momentum 4.

\subsection{Effect of hybridization between $j=\f32$ and $j=\f12$ states\lbl{sec:KMhybr}}

Our main interest is the behavior of $j=\f32$ states at energies $|\e|\ll|\De|$ close to the $j=\f32$ level $\e=0$.
Exactly at $\pb=\nv$, the $j=\f32$ and $j=\f12$ states are decoupled.
However, even at small momenta hybridization to $j=\f12$ states affects the properties of $j=\f32$ states.
Therefore, simply neglecting the hybridization $\Hh_{\f12\f32}(\pb)$ to $j=\f12$ states in the KM \eqn{HK}
and considering the block $\Hh_{\f32\f32}(\pb)=\Hh^\Lx(\pb)|_{\al_{0,z,\square}=\ga_{0,z,\square}}$
with ``bare'' parameters $\ga_{0,z,\square}$ as the Hamiltonian for $j=\f32$ states would be incorrect.

The effect of hybridization is best illustrated by considering momenta $\pb=(0,0,p_z)$.
For clarity, we also consider the case of full spherical symmetry $\Ox(3)$ with inversion, putting $\ga_\square=0$;
the corresponding quantities will be labeled with $\Ox(3)$ superscript.
The $\Ox(3)$-symmetric Kane Hamiltonian $\Hh^{\Kx,\Ox(3)}(0,0,p_z)\equiv \Hh^\Kx(0,0,p_z)|_{\ga_\square=0}$
at a given $p_z\neq0$ possesses axial symmetry
with respect to rotations about the $z$ axis and the states with different $j_z$ are decoupled.

The $j_z=\pm\f12$ states are present for both $j=\f32$ and $j=\f12$
and there is hybridization between them.
For both pairs $(\Psi_{\f12,\pm\f12},\Psi_{\f32,\pm\f12})$,
the $2\tm 2$ Hamiltonian has the form
\beq
    \Hh_{|j_z|=\f12}^{\Ox(3)}(0,0,p_z)
    =\lt(\ba{cc} \De+\ga_{\f12}p_z^2 & v\sq{\f23}p_z \\ v\sq{\f23}p_z & (\ga_0+2\ga_z)p_z^2 \ea\rt).
\lbl{eq:Hpzf12}
\eeq
Due to $\Ox(3)$ symmetry, the spectrum is isotropic; so,
diagonalizing \eq{Hpzf12} and replacing $p_z^2\rarr p^2\equiv\pb^2=p_x^2+p_y^2+p_z^2$,
we get two double-degenerate bands
\beq
    \eps^{\Kx,\Ox(3)}_{a,b}(p)=\f12\lt\{
    \De+(\ga_{\f12}+\ga_0+2\ga_z)p^2\pm\sq{
    [-\De+(-\ga_{\f12}+\ga_0+2\ga_z)p^2]^2+\f83 v^2p^2}
    \rt\}.
\lbl{eq:eKab}
\eeq
At small momenta, the band originating from the $j=\f32$ level $\e=0$ at $p=0$ has the form
\beq
    \eps^{\Kx,\Ox(3)}_a(p)=(\ga_0+2\ga_z-\f23\f{v^2}{\De})p^2+\Oc(p^4)
\lbl{eq:eKexp}
\eeq
We see that, indeed, due to hybridization,
the spectrum is modified compared to the respective band $(\ga_0+2\ga_z)p^2$
of the $\Hh_{\f32\f32}(\pb)|_{\ga_\square=0}$ block with bare parameters $\ga_{0,z}$.

On the other hand, the $j_z=\pm\f32$ states are present only for $j=\f32$ and thus
they do not hybridize to $j=\f12$ states.
For both $\Psi_{\f32,\pm\f32}$, the scalar Hamiltonian reads
\beq
    \Hh^{\Ox(3)}_{|j_z|=\f32}(0,0,p_z)
    =(\ga_0-2\ga_z)p_z^2.
\lbl{eq:Hpzf32}
\eeq
It gives one double-degenerate band
\beq
    \eps^{\Kx,\Ox(3)}_c(p)=(\ga_0-2\ga_z)p^2,
\lbl{eq:eKc}
\eeq
exactly equal to that of the $\Hh_{\f32\f32}(\pb)|_{\ga_\square=0}$ block.


\subsection{Folding procedure, effective Hamiltonian for the Luttinger model for $j=\f32$ states \lbl{sec:fpH}}

To account for the effect of hybridization, a systematic ``folding'' procedure~\cite{WS}
must be performed for both the bulk Hamiltonian and BCs,
where the high-energy $j=\f12$ states are consistently eliminated from the Hilbert space,
while the effect of virtual transitions to them is taken into account.

For the bulk Hamiltonian, the procedure is as follows.
Excluding $\Psih_{\f12}$ [\eq{Psi}] from the Schr\"odinger equation $\Hh^\Kx(\pb)\Psih=\e\Psih$,
we obtain the equation
\beq
    \lt(\Hh_{\f32\f32}(\pb)+\Hh_{\f32\f12}(\pb)\f1{\e\um_2-\Hh_{\f12\f12}(\pb)}\Hh_{\f12\f32}(\pb)\rt)\Psih_{\f32}=\e\Psih_{\f32}
\lbl{eq:scheq32}
\eeq
for $\Psih_{\f32}$.
At $|\e|\ll|\De|$ and $\ga_{\f12} p^2 \ll |\De|$,
the energy $\e$ and momentum $\pb$ should be set to zero in the denominator in the left-hand side.
After that, \eq{scheq32} becomes an effective Schr\"odinger equation
\[
    \Hh^\Lx(\pb)\psih^\Lx=\e\psih^\Lx
\]
for $j=\f32$ states only with the 4-component wave function
\beq
    \psih^\Lx=\lt(\ba{c} \psi^\Lx_{+\f32} \\ \psi^\Lx_{+\f12} \\ \psi^\Lx_{-\f12} \\ \psi^\Lx_{-\f32}\ea\rt),
\lbl{eq:psiL}
\eeq
for which
\beq
    \Psih_{\f32}\rarr \psih^\Lx
\lbl{eq:Psi32->psiL}
\eeq
needs to be substituted.

Expectedly, the effective Hamiltonian
\beq
    \Hh^\Lx(\pb)=
    \Hh_{\f32\f32}(\pb)
    +\Hh_{\f32\f12}(\pb)\f1{0\um_2-\Hh_{\f12\f12}(\nv)}\Hh_{\f12\f32}(\pb)
\lbl{eq:HLeff}
\eeq
has the form
(13)
of the LM with parameters
\[
    \al_0=\ga_0-\f{v^2}{3\De}
,\spc
    \al_z=\ga_z-\f{v^2}{6\De}
,\spc
    \al_\square=\ga_\square.
\]
The parameters $\al_{0,z}$ of the $\Ox(3)$-symmetric part of the LM are modified,
while the cubic anisotropy parameter $\al_\square$ is not,
since the terms involved in the hybridization have $\Ox(3)$ symmetry.

The spectrum of the $\Ox(3)$-symmetric LM $\Hh^{\Lx,\Ox(3)}(\pb)\equiv \Hh^\Lx(\pb)|_{\ga_\square=0}$
for $j=\f32$ states following from the KM
consists of two double-degenerate bands
\beq
    \eps^{\Lx,\Ox(3)}_+(p)=(\al_0+2\al_z)p^2=(\ga_0+2\ga_z-\f23\f{v^2}{\De})p^2,
\lbl{eq:eL+}
\eeq
\beq
    \eps^{\Lx,\Ox(3)}_-(p)=(\al_0-2\al_z)p^2=(\ga_0-2\ga_z)p^2.
\lbl{eq:eL-}
\eeq
They agree with the bands \eqsn{eKexp}{eKc}, respectively, of the $\Ox(3)$-symmetric KM
at small momenta, where the former band is affected by hybridization with $j=\f12$ states and the latter is not.

For many semiconductor materials, the bare parameters are such that $\ga_{\f12}>0$, $\ga_{0,z}<0$, $\ga_0-2\ga_z<0$,
and, neglecting hybridization to $j=\f12$ states ($v=0$),
both bands $(\ga_0\pm2\ga_z)p^2<0$ of $j=\f32$ states
would be hole-like.
In the noninverted regime of the KM, $\De>0$ and the $j=\f12$ level $\e=\De$ is above the $j=\f32$ level $\e=0$.
In this case, the system is an insulator and the band $\eps^{\Lx,\Ox(3)}_+(p)$ [\eq{eL+}]
is pushed further down by hybridization.
In the inverted regime of the KM, $\De<0$ and the $j=\f12$ level $\e=-|\De|$ is below the $j=\f32$ level $\e=0$,
and the band $\eps^{\Lx,\Ox(3)}_+(p)$ is pulled up by hybridization.
For weaker hybridization, $\ga_0+2\ga_z+\f23\f{v^2}{|\De|}<0$,
the band $\eps^{\Lx,\Ox(3)}_+(p)$ is still hole-like, and there is also a Fermi surface at $\e=0$.
For stronger hybridization, such that $\ga_0+2\ga_z+\f23\f{v^2}{|\De|}>0$,
the band $\eps^{\Lx,\Ox(3)}_+(p)$ is electron-like, and the system is a nodal semimetal.
This regime is realized in $\al$-Sn, HgTe and many similar materials.

\subsection{Folding procedure, effective boundary conditions for the Luttinger model for $j=\f32$ states \lbl{sec:fpbc}}

Considering the more general KM allows us to derive asymptotic BCs for the LM wave function $\psih^\Lx(x,y,z)$ [\eq{psiL}].
We derive the BCs explicitly for $\Ox(3)$ and argue below that they holds for cubic symmetry $\Og_h$ as well.
Due to $\Ox(3)$ symmetry, it is sufficient to consider any surface orientation; we choose the $z=0$ surface
and assume the sample occupies the $z>0$ half-plane.

We consider the so-called ``hard-wall'' BCs for the KM,
\beq
    \Psih(x,y,z=0)=\nm,
\lbl{eq:KMbc}
\eeq
for which the wave function vanishes at the surface.
Such BCs represent an interface with vacuum,
which can be described by the KM \eqn{HK} in the noninverted regime (trivial insulator) with infinite gap $\De\rarr+\iy$.

As for the effective bulk Hamiltonian, a ``folding procedure'' must be carried for BCs.
The general idea (also applicable to other similar situations)
of deriving the asymptotic BCs for the low-energy LM
from a more general KM with a larger Hilbert space is as follows.
We look for the general solution to the Scr\"odinger equation
\beq
    \Hh^{\Kx,\Ox(3)}(0,0,\ph_z)\Psih(z)=\nm
\lbl{eq:HKpzSE}
\eeq
($\ph_z=-\ix\pd_z$ is the momentum operator)
for the KM exactly at energy $\e=0$ of the $j=\f32$ states at $\pb=\nv$.
The wave function $\Psih(z)$ depends only on $z$ in this case.

Partial solutions to \eq{HKpzSE} are described by
the momentum solutions $p_z$ to its characteristic equation $\det\Hh^{\Kx,\Ox(3)}(0,0,p_z)=0$.
Since the energy $\e=0$ taken is right at the node,
either $p_z=0$ or $p_z$ contain imaginary parts.
For $p_z=0$, the wave function contains linear polynomials, $\Psih(z)\larr 1,z$.
These represent the low-energy part of the solution,
which should be identified with the LM wave function \eqn{psiL}
by matching with the first two terms of its Taylor expansion
\beq
    \psih^\Lx(x,y,z)=\psih^\Lx(x,y,0)+\pd_z\psih^\Lx(x,y,0)z+\Oc(z^2)
\lbl{eq:psiLexp}
\eeq
at the surface.

For $p_z$ with imaginary parts, partial solutions are exponentials, $\Psih(z)\larr\ex^{\ix p_z z}$,
that decay or grow into the bulk. The growing exponentials must be discarded, while the decaying ones retained.
Imposing the BCs \eqn{KMbc} on the solution $\Psih(z)$
that is a linear combination of the low-energy part with $p_z=0$
and exponentially decaying solutions and eliminating the latter,
one arrives at the BCs for the LM wave function $\psih^\Lx$.

Below we carry out this procedure explicitly.
The convenience of considering the $z=0$ surface and $\Ox(3)$ symmetry is that,
as explained above, due to axial symmetry,
the Hamiltonian $\Hh^{\Kx,\Ox(3)}(0,0,\ph_z)$ is decoupled
into the blocks \eqsn{Hpzf12}{Hpzf32} for states with given $j_z$.

For the pairs $(\Psi_{\f12,\pm\f12}(z),\Psi_{\f32,\pm\f12}(z))$,
the characteristic equation reads
\beq
    \det\Hh^{\Ox(3)}_{|j_z|=\f12}(0,0,p_z)=
    p_z^2[(\ga_0+2\ga_z)(\De+\ga_{\f12}p_z^2)-\tf23v^2]=0.
\lbl{eq:detHpzf12=0}
\eeq
There is a doubly-degenerate solution $p_z=0$.
The corresponding partial solutions to
\beq
    \Hh^{\Ox(3)}_{|j_z|=\f12}(0,0,\ph_z)
        \lt(\ba{c}\Psi_{\f12,\pm\f12}(z) \\ \Psi_{\f32,\pm\f12}(z)\ea\rt)=\nm
\lbl{eq:Hpzf12SE}
\eeq
are
\[
    \lt(\ba{c}0\\1\ea\rt)
    ,\spc
    \lt(\ba{c} \ix \sq{\f23} \f{v}{\De}\\z\ea\rt).
\]

The other two solutions
\beq
    p_z=\pm\ix\ka, \spc \ka=\sq{\f{\De}{\ga_{\f12}}\f{\al_0+2\al_z}{\ga_0+2\ga_z}},
\lbl{eq:ka}
\eeq
to \eq{detHpzf12=0} are purely imaginary.
The partial solution
\[
    \lt(\ba{c} \Psir_{\f12,\pm\f12} \\ \Psir_{\f32,\pm\f12} \ea\rt)
    \ex^{-\ka z}
    = \lt(\ba{c} \ga_0+2\ga_z \\ \ix\sq{\f23}\f{v}{\ka}  \ea\rt)
    \ex^{-\ka z}
\]
to \eq{Hpzf12SE} with $p_z=\ix\ka$ decays into the bulk and is admitted,
while the partial solution with $p_z=-\ix\ka$ grows into the bulk and is prohibited.

Altogether, the general solution to \eq{Hpzf12SE}, not growing exponentially into the bulk,
is the linear combination
\beq
    \lt(\ba{c} \Psi_{\f12,\pm\f12}(z) \\ \Psi_{\f32,\pm\f12}(z) \ea\rt)
    =\psi_{\pm\f12}^\Lx\lt(\ba{c}0\\1\ea\rt)
    +\pd_z\psi_{\pm\f12}^\Lx\lt(\ba{c} \ix\sq{\f23}\f{v}{\De} \\z\ea\rt)
    +C\lt(\ba{c} \Psir_{\f12,\pm\f12} \\ \Psir_{\f32,\pm\f12} \ea\rt)\ex^{-\ka z}
\lbl{eq:Psif12bc}
\eeq
with three {\em constant} coefficients $\psi_{\pm\f12}^\Lx$, $\pd_z\psi_{\pm\f12}^\Lx$, and $C$.

The identification of these coefficients
with the LM wave function $\psih^\Lx(x,y,z)$ [\eq{psiL}] is performed as follows.
On the one hand, at distances $\ka z\gg 1$, the exponential partial solution in \eq{Psif12bc}
has decayed and
\beq
    \Psi_{\f32,\pm\f12}(z)
    = \psi_{\pm\f12}^\Lx+\pd_z\psi_{\pm\f12}^\Lx z+\Oc(\ex^{-\ka z}).
\lbl{eq:Psi3212exp}
\eeq
One the other hand, $\psih^\Lx(x,y,z)$
varies over spatial scales much larger than $1/\ka$, which is set by $\De$ [\eq{ka}].
At intermediate spatial scales, both approximate forms \eqn{psiLexp} and \eqn{Psi3212exp}
are valid, and, according to the correspondence \eqn{Psi32->psiL}, the constants
\beqarn
    \psi_{\pm\f12}^\Lx&\rarr&\psi_{\pm\f12}^\Lx(x,y,0), \lbl{eq:}\\
    \pd_z\psi_{\pm\f12}^\Lx&\rarr&\pd_z\psi_{\pm\f12}^\Lx(x,y,0) \lbl{eq:}
\eeqarn
need to be identified with the components of the LM wave function \eqn{psiL} at the surface and their first derivatives.

Imposing the hard-wall BCs \eqn{KMbc} on \eq{Psif12bc} gives
\[
    \lt\{\ba{l}
        \pd_z\psi_{\pm\f12}^\Lx \ix\sq{\tf23}\f{v}{\De}+C\Psir_{\f12,+\f12}=0,\\
        \psi_{\pm\f12}^\Lx+C\Psir_{\f32,+\f12}=0.
    \ea\rt.
\]
Excluding $C$, we arrive at the constraint
\[
    \psi^\Lx_{\pm\f12}+l_\De\pd_z\psi^\Lx_{\pm\f12}=0,
\]
where
\[
    l_\De=\f23\f{v^2}{\De\ka(\ga_0+2\ga_z)}
\]
is a spatial scale set by $\De$. Since $\psih^\Lx(x,y,z)$ varies over larger scales,
the second term with the derivative must be neglected
(keeping it would be exceeding the accuracy;
this explicitly illustrates the point about long-wavelength limit and {\em asymptotic} BCs made in the Main Text.)
and we arrive at the asymptotic BCs
\[
    \psi^\Lx_{\pm\f12}(x,y,0)=0.
\]

For the $\Psi_{\f32,\pm\f32}(z)$ components, not coupled to $j=\f12$ states, the analogous procedure is trivial.
The momentum solution $p_z=0$ to the characteristic equation
\[
    \det\Hh^{\Ox(3)}_{|j_z|=\f32}(0,0,p_z)=(\ga_0-2\ga_z)p_z^2=0
\]
is double-degenerate and the corresponding solution to
\[
    \Hh^{\Ox(3)}_{|j_z|=\f32}(0,0,\ph_z)\Psi_{\f32,\pm\f32}(z)=0
\]
reads
\[
    \Psi_{\f32,\pm\f32}(z)=\psi_{\pm\f32}^\Lx+\pd_z\psi_{\pm\f32}^\Lx z.
\]
Imposing the hard-wall BCs \eqn{KMbc}, upon the identification
\beqarn
    \psi_{\pm\f32}^\Lx&\rarr&\psi_{\pm\f32}^\Lx(x,y,0), \lbl{eq:}\\
    \pd_z\psi_{\pm\f32}^\Lx&\rarr&\pd_z\psi_{\pm\f32}^\Lx(x,y,0), \lbl{eq:}
\eeqarn
we obtain the asymptotic BCs
\[
    \psi_{\pm\f32}^\Lx(x,y,0)=0.
\]

And so, the asymptotic BCs for the LM wave function $\psih^\Lx$ [\eq{psiL}]
corresponding to the hard-wall BCs \eqn{KMbc}
for the KM wave function $\Psih$ [\eq{Psi}] are given by
Eq.~(14),
with all components vanishing at the boundary.

Clearly, for $\Ox(3)$ symmetry, the BCs
(14)
are valid for any orientation of the surface
(one can use the angular-momentum basis with the quantization axis perpendicular to that surface).
Moreover, these BCs also hold when the symmetry is lowered to cubic $\Og_h$,
since this form of BCs with all four wave-function components vanishing
cannot be continuously transformed to any other possible form of asymptotic BCs,
which would necessarily also involve first-order derivatives.

Regarding the physical meaning, we caution from interpreting the BCs
(14)
as the ``hard-wall'' BCs,
since such interpretation implies an infinite potential barrier, which is not meaningful in the semimetal regime of the LM.